\documentclass[12pt,notitlepage]{article}
\usepackage{amssymb}

\usepackage{graphicx}
\usepackage{amsmath}

\input{tcilatex}

\begin{document}

\begin{center}
{\LARGE SPACE-TIME UNCERTAINTY FROM HIGHER-DIMENSIONAL DETERMINISM}

\bigskip

\bigskip

\bigskip 

{\Large (Or: How Heisenberg was right in 4D because Einstein was right in 5D)%
}

\bigskip

\bigskip 
Paul S. Wesson$^{1,2}$

\bigskip

\bigskip 

\bigskip 
\end{center}

\begin{enumerate}
\item Dept. of Physics, University of Waterloo, Waterloo, Ontario \ \ N2L 3G1

\item Space-Time-Matter Group Webpage: http://astro.uwaterloo.ca/\symbol{126}%
wesson)
\end{enumerate}



Correspondence: post to address (1) above, phone (519)885-1211 x2215, fax
(519)746-8115, email wesson@astro.uwaterloo.ca

\pagebreak

\bigskip \underline{{\Large Abstract}}

Heisenberg's uncertainty relation is commonly regarded as defining a level
of unpredictability that is fundamentally incompatible with the
deterministic laws embodied in classical field theories such as Einstein's
general relativity. \ We here show that this is not necessarily the case. \
Using 5D as an example of dimensionally-extended relativity, we employ a
novel metric to derive the standard quantum rule for the action and a form
of Heisenberg's relation that applies to real and virtual particles. \ The
philosophical implications of these technical results are somewhat profound.

\begin{center}
*
\end{center}

Einstein and Heisenberg espoused fundamentally different views of mechanics,
the former arguing that God does not play dice with the world, and the
latter seeing indeterminacy as an essential aspect of it. \ This problem has
recently come to the fore again as part of new attempts to unify gravity
with the interactions of particle physics. \ The best route to unification
is commonly regarded as dimensional extension, wherein four-dimensional (4D)
spacetime is augmented by extra parts$^{1}$. \ Hence 5D induced-matter and
membrane theory (both modern versions of Kaluza-Klein theory), 10D
supersymmetry, 11D supergravity and 26D string theory. \ Does the algebraic
richness of these new theories offer a way out of the old physical problem?

Mechanics is often regarded as a staid subject, but it is actually full of
conceptual twists that offer clues to how we might answer the above question$%
^{2}$. \ Newton's mechanics defines force as the product of mass and
acceleration, where the latter is calculated from separate space $\left(
x\right) $ and $\left( t\right) $ coordinates, while the former is just
given. \ Mach's mechanics postulates that the local (inertial rest) mass $m$
of a particle ought to be calculable as the sum of other influences in the
universe; but while this principle motivated Einstein, it was not properly
incorporated into general relativity. \ The latter is a curved-space version
of special relativity, whose underlying framework is due to Minkowski. \ He
realized that the speed of light allows the time to be treated as a
coordinate that is on the same footing as those of space $\left( x^{\alpha
}=ct,x,y,z\right) $. \ Both versions of relativity are classical field
theories, built along the lines of Maxwell's theory of electromagnetism, and
neither incorporates at a basic level the quantum of action $h$ named after
Planck. \ Indeed, an examination of the foundations of classical and quantum
mechanics shows that the former is a theory of accelerations while the
latter is a theory of changes in momentum or forces. \ The two concepts
overlap, of course, but are only equivalent when the mass is constant. \
Examples where this is not so involve the acceleration of a rocket as it
burns fuel and leaves the Earth,$^{2}$ or the process by which particles
gain mass by a Higgs-type scalar field in the early universe.$^{3}$ \ This
difference has been commented on by a number of astute workers, and reflects
our lack of a theory of the origin of mass.$^{4}$ \ This is pointed up by
the fact that there is an ambiguity between the mass of a particle and the
energy of a resonance in particle interactions, and that both are related
via a Heisenberg-type relation with their lifetimes. \ In relativity as
applied to either classical or quantum situations, the ordinary time $t$ is
replaced by the proper time $\tau $ of a flat Minkowski metric or the
interval $s$ of a curved Riemannian metric. \ The latter is defined by $%
ds^{2}=g_{\alpha \beta }dx^{\alpha }dx^{\beta }$. \ (Here $g_{\alpha \beta }$
are the components of the metric tensor, which are basically gravitational
potentials, and there is a sum over the repeated indices $\alpha =0,123$.) \
There are two equivalent ways to obtain the dynamics of a particle in such a
metric-based theory. \ One involves forming the path length $\left( \int
ds\right) $, and using a version of Fermat's theorem to minimize this $%
\left( \delta \left[ \int ds\right] =0\right) $, which results in the four
components of the geodesic equation. \ The other way is to form a Lagrangian
density from $s$ or a function of it, employ the Euler-Lagrange equations,
and so obtain the four equations of motion. \ [In both approaches, the
energy of a particle and its momentum in ordinary space are associated with
the $t,\,xyz$ or 0, 123 components.] \ However, these approaches necessarily
result in equations for the acceleration, because the mass $m$ is absent. \
By contrast, the action of 4D particle physics $\left( \int mcds\right) $
involves the mass explicitly. \ This mismatch in the ways in which dynamics
is set up becomes acute when we attempt to extend the manifold in such a way
as to account for the mass-insensitive nature of gravity and the
mass-sensitive nature of particle interactions.

Fortunately, we are aided by some new and startling results from 5D
relativity. \ There are two popular versions of this. \ Induced-matter
theory views mass as a direct manifestation of an unconstrained fifth
dimension, and its corpus of technical results is largely based on the
canonical metric.$^{5,6}$ \ Membrane theory views matter as confined to a
surface in a 5D world, and most of its technical results are based on the
warp metric.$^{7,8}$ \ Both theories regard 4D spacetime as embedded in a 5D
manifold whose line element is defined by $dS^{2}=g_{AB}dx^{A}dx^{B}$ ($A,B=$%
0, 123, 4 where the extra coordinate will henceforth be labelled $x^{4}=l$).
\ Both theories are well-regarded because they represent the basic extension
of general relativity and the low-energy limit of higher-$N$ accounts. \ In
addition, it has recently been shown that the field equations of these
theories are in fact equivalent, so their solutions for the potentials $%
g_{AB}$ are common.$^{9}$ \ We will not be concerned with the field
equations in what follows, but mention two results to do with dynamics that
are also common: (a) Massive particles travelling on timelike geodesics in
4D $\left( ds^{2}>0\right) $ can be regarded as travelling on \underline{null%
} geodesics in 5D $\left( dS^{2}=0\right) .^{10,11}$ \ In other words, what
we regard as ordinary objects in spacetime are like photons in the bigger
manifold, and \underline{they are in causal contact}. \ This has obvious
implications for the wave-like behaviour of particles, such as those of
electrons in the classic double-slit experiment. \ (b) Massive particles
travelling along $s$-paths in 4D in general change their mass via $m=m\left(
s\right) $, and this is connected to the existence of a fifth force which is
due to the fifth dimension.$^{12,13}$ \ That such a new force exists can be
readily seen, by noting that in theories like general relativity there is an
orthogonality condition which relates the 4-velocities $\left( u^{\alpha
}\equiv dx^{\alpha }/ds\right) $ to the components of the acceleration or
force per unit mass $\left( f_{\alpha }\right) $, namely $u^{\alpha
}f_{\alpha }=0.$ \ But in 5D, the corresponding relation is $u^{A}f_{A}=0$,
\ so perforce $u^{\alpha }f_{\alpha }=-u^{4}f_{4}\neq 0$. \ Also, the new
force acts \underline{parallel} to the 4-velocity. \ Provided we use the
proper time $s$ to parametrize the motion, and seek to make contact with the
large amount of data we already have on 4D dynamics, the logical way to
quantify this (presumably small) force is via a variation in the mass. \
[Depending on whether one uses the canonical or warp metric, the mass itself
is either the coordinate or its rate of change, but these two
identifications are essentially equivalent for null 5D geodesics, as will be
noted below.] \ A force which acts parallel to the 4-velocity, and changes
the mass and therefore the momentum, is new to classical field theory but
not to quantum theory. \ Is there a link? \ We conjecture that $u^{\alpha
}f_{\alpha }\neq 0$ is related to $dx^{\alpha }dp_{\alpha }\neq 0$; and that
the classical laws of an extended manifold are related to the quantum
uncertainty of spacetime.

We now proceed to show, in short order, how to derive Heisenberg's
uncertainty relation in 4D from Einstein-like deterministic dynamics in 5D.
\ We will use the Lagrangian approach, but the results are compatible with a
longer approach based on the geodesic equation$^{14,15}$. \ Conventional
units are retained for ease of physical interpretation. \ We will use the
notation of induced-matter theory as opposed to membrane theory, because it
is more direct. \ However, we introduce \ a new form for the line element,
which for reasons that will become apparent we refer to as the Planck gauge:%
\begin{equation}
dS^{2}=\frac{L^{2}}{l^{2}}\,g_{\alpha \beta }\left( x^{\gamma },l\right)
dx^{\alpha }dx^{\beta }-\frac{L^{4}}{l^{4}}dl^{2}\;\;\;\;.
\end{equation}%
Here, we have used 4 of the 5 available degrees of coordinate freedom to
remove the $g_{4\alpha }$ components of the metric that are traditionally
associated with electromagnetism; and we have used the fifth degree of
coordinate freedom to set $g_{44}=-L^{4}\diagup l^{4}$. \ Algebraically, (1)
is therefore general, insofar as its 4D part has been factorized by $%
L^{2}\diagup l^{2}$ but $g_{\alpha \beta }=g_{\alpha \beta }\left( x^{\gamma
},l\right) $ is still free. \ Physically, however, we do not expect any
incursion of the fifth dimension into 4D spacetime, since this would violate
the Weak Equivalence Principle$^{1,2}$. \ We therefore proceed with $%
g_{\alpha \beta }=g_{\alpha \beta }\left( x^{\gamma }\right) $. \ Then, it
may be mentioned in passing that the coordinate transformation (change of
gauge) $l\rightarrow L^{2}\diagup l$ converts (1) to the canonical metric of
induced-matter theory$^{5,6}$, and that a further transformation converts it
to the warp metric of membrane theory.$^{7,8}$ \ The constant $L$ in (1) is
required by dimensional consistency, and corresponds physically to the
characteristic size of the potential in the fifth dimension. \ However, it
also sets a scale for 4D spacetime. \ [For the whole universe, a reduction
of the field equations shows that $L=\left( 3\diagup \Lambda \right) ^{1/2}$
where $\Lambda $\ is the cosmological constant.$^{1}$] \ From (1), we can
form the Lagrangian density $\left( dS\diagup ds\right) ^{2}$ and use the
Euler-Lagrange equations to obtain the associated 5-momenta in the standard
manner. \ These 5-momenta define a 5D scalar which is the analog of the one
used in 4D quantum mechanics. \ Following the philosophy outlined above, we
split it into its 4D and fifth parts as follows:%
\begin{equation}
\int P_{A}\;dx^{A}=\int \left( P_{\alpha }dx^{\alpha }+P_{l}\;dl\right)
=\int \frac{2L^{2}}{l^{2}}\left[ 1-\left( \frac{L}{l}\,\frac{dl}{ds}\right)
^{2}\right] ds\;\;\;\;.
\end{equation}%
However, this is actually zero for particles moving on \underline{null}
geodesics of (1) with $dS^{2}=0$, as for induced-matter and membrane theory.
\ Then $l=l_{0}\exp \left( \pm s\diagup L\right) $ and $dl\diagup ds=\pm
\left( l\diagup L\right) $, so the variation is slow for $s\diagup L\ll 1$
and the mass parametrizations for the two theories are equivalent modulo $L$%
, as noted above. \ For (1), the appropriate parametrization is clearly $%
l=h\diagup mc$, which means that the extra coordinate is the Compton
wavelength of the particle. \ The first part of (1) then gives the
conventional action of 4D particle physics. \ But note the important
difference that while the 4D action is finite and describes a particle with
finite energy, the 5D action is zero because $\int P_{A}\,dx^{A}=0$.

The quantity that corresponds to this in 4D is%
\begin{equation}
\int p_{\alpha }dx^{\alpha }=\int mu_{\alpha }dx^{\alpha }=\int \frac{h\,ds}{%
cl}=\pm \frac{h}{c}\,\frac{L}{l}\;\;\;\;.
\end{equation}%
The sign choice here goes back to the reversible nature of the motion in the
fifth dimension, but this is unimportant so we suppress it in what follows.
\ We also put $L\diagup l=n$, to make contact with other work on the wave
nature of particles. \ Then (3) says%
\begin{equation}
\int mcds=nh\;\;\;\;,
\end{equation}%
which of course is known to every physics student. \ More revealingly, from
previous relations we can form the scalar%
\begin{equation}
dp_{\alpha }dx^{\alpha }=\frac{h}{c}\left( \frac{du_{\alpha }}{ds}\frac{%
dx^{\alpha }}{ds}-\frac{1}{l}\frac{dl}{ds}\right) \frac{ds^{2}}{l}\;\;\;\;.
\end{equation}%
Here the first term is conventional, and is zero if the acceleration is zero
and / or if the standard orthogonality relation $f_{\alpha }u^{\alpha }=0$
holds (see above). \ But the second term is unconventional, and is in
general non-zero. \ It measures the variation in the (inertial rest) mass of
a particle required to balance the extended laws of conservation. \ The
anomalous contribution has magnitude%
\begin{equation}
\left| dp_{\alpha \,}dx^{\alpha }\right| =\frac{h}{c}\left| \frac{dl}{ds}%
\right| \frac{ds^{2}}{l^{2}}=\frac{h}{c}\frac{ds^{2}}{Ll}=n\frac{h}{c}\left( 
\frac{dl}{l}\right) ^{2}\;\;\;\;.
\end{equation}%
This is a Heisenberg-type relation. \ Reading it from right to left, it says
that the fractional change in the position of the particle in the fifth
dimension has to be matched by a change of its dynamical quantities in 4D
spacetime. \ We can write (6) in more familiar form as 
\begin{equation}
\left| dp_{\alpha }dx^{\alpha }\right| =\frac{h}{c}\frac{dn^{2}}{n}\;\;\;\;.
\end{equation}%
This is as far as the formal analysis goes. \ We have not so far considered
questions of topology, but there are clearly two cases of (7) which depend
on this. \ For a particle trapped in a potential box, the Compton wavelength
cannot exceed the confining size of the geometry, so $l\leq L$, $n\geq 1$
and we have a violation of the Heisenberg rule as for virtual particles. \
For a particle that is free, the Compton wavelength is unconstrained, so $%
l>L $, $n<1$ and we have the conventional Heisenberg rule as for real
particles.

In conclusion, we review our technical results and make some philosophical
comments. \ In establishing a consistent scheme for dynamics in more than 4
dimensions, we have been guided by the history of mechanics and have chosen
to make contact with extant results by using the proper 4D time as
parameter. \ Of the infinite number of metrics or gauges available, we have
proposed that (1) is the most appropriate for applications to particle
physics. \ Inspection shows that this works well because it incorporates the
(inertial rest) mass as a coordinate in such a way as to convert the metric
to a momentum manifold. \ This allows the theory of accelerations as used in
classical field theories like general relativity to be extended to a theory
of momentum changes (forces) as used in quantum theory. \ The main results
are the standard quantization rule for the action (4) and a Heisenberg-type
uncertainty relation (7). \ An insight from the latter is that virtual and
real particles are aspects of the same underlying dynamics, separated by a
number $0<n<\infty $ at $n=1$. \ These technical results are neat (and
clearly beg for more investigation); but the philosophical implications of
the preceding outline are more profound.

Since the 1930s, the view has become ingrained that quantum physics
necessarily involves a level of uncertainty or non-predictability that is in
conflict with the deterministic laws of classical field theory as
represented by Maxwell's electromagnetism and Einstein's general relativity.
\ Surprisingly, we now see that this is not necessarily the case. \ The
dichotomy may be an artificial one, laid on us by our ignorance of the
extent of the real world. \ In a classical field theory like general
relativity which is extended to $N(>4)$ dimensions, the extra parts of
deterministic laws can manifest themselves in spacetime as ``anomalous''
effects. \ This will perforce happen in any $ND$ field theory that is not
artificially constrained by ``symmetries''. \ The last word has,
unfortunately, been used all too often as an excuse for algebraic shortcuts
that are not respected by reality. \ A prime example is the ``cylinder
condition'' of old 5D Kaluza-Klein theory. \ By removing dependency on the
extra coordinate, \ it emasculates the metric and runs the theory into
insurmountable difficulties to do with the masses of elementary particles
and the energy density of the vacuum (the hierarchy and
cosmological-constant problems). \ The modern versions of 5D relativity,
namely induced-matter theory and membrane theory, drop the cylinder
condition. \ These theories may be daunting algebraically, but they gain in
being richer physically. \ The same lesson can be applied to 10D
supersymmetry, 11D supergravity and 26D string theory. \ We should recall,
in this regard, that Einstein in his later work endorsed higher dimensions
and remained adamant that quantum uncertainty was philosophically
unacceptable. \ The results we have shown here could provisionally be used
to paraphrase Einstein: ``God does not play dice in a higher-dimensional
world.''

\bigskip

\underline{{\Large Acknowledgements}}

The work reported here is based on previous work and comments by members of
the Space-Time-Matter Consortium. \ It was supported by N.S.E.R.C.

\bigskip

\underline{{\Large References}}

\begin{enumerate}
\item Wesson, P.S. Space, Time, Matter (World Scientific, Singapore, 1999).

\item Rindler, W. Essential Relativity (Springer, New York, 1977).

\item Linde, A.D. Inflation and Quantum Cosmology (Academic, Boston, 1990).

\item Jammer, M. Concepts of Mass in Contemporary Physics and Philosophy
(Princeton Un. Press, Princeton, 2000).

\item Wesson, P. Phys. Lett. B\underline{276}, 299 (1992).

\item Mashhoon, B., Wesson, P.S., Liu, H. Gen. Rel. Grav. \underline{30},
555 (1998).

\item Randall, L., Sundrum, R. Mod. Phys. Lett. A\underline{13}, 2807 (1998).

\item Arkani-Hamed, N., Dimopoulous, S., Dvali, G.R. Phys. Lett. B\underline{%
429}, 263 (1998).

\item Ponce de Leon, J. Mod. Phys. Lett. A\underline{16}, 2291 (2001).

\item Seahra, S.S., Wesson, P.S. Gen. Rel. Grav. \underline{33}, 1731 (2001).

\item Youm, D. Mod. Phys. Lett. A\underline{16}, 2371 (2001).

\item Wesson, P.S. Mashhoon, B., Liu, H., Sajko, W.N. Phys. Lett. B%
\underline{456}, 34 (1999).

\item Youm, D. Phys. Rev. D\underline{62}, 084002 (2000).

\item Wesson, P.S. J. Math. Phys. \underline{43}, 2423 (2002).

\item Wesson, P.S. Class. Quant. Grav. \underline{19}, 2825 (2002).
\end{enumerate}

\end{document}